\documentclass[10pt,twocolumn,a4paper]{article}
\usepackage{amsmath,amsfonts,amssymb,amstext}
\usepackage[latin1]{inputenc}
\usepackage{rawfonts,verbatim,color}
\usepackage{graphicx}         
\usepackage{dcolumn,multirow}
\usepackage[margin=15pt,font=small,labelfont=bf,labelsep=period]{caption}

\mathchardef\mhyphen="2D
\newcommand{\nitem}{\vspace{-0.4em}\item}
\newcommand{\nin}{\noindent}
\newcommand{\Mol}{M\o{}ller}

\newcommand{\rmd}{\mathrm{d}}

\newcommand{\ee}{\mbox{$e^-$--$e^-$}}
\newcommand{\ep}{$e^-$--$e^+${}}
\newcommand{\sigobs}{\sigma_{\sf obs}^{\prime}}
\newcommand{\spr}{\sigma^{\prime}}

\newcommand{\Es}{E_{\sf s}}
\newcommand{\xs}{x_{\sf s}}
\newcommand{\ths}{\theta_{\sf s}}
\newcommand{\adag}{a^{\dag}}

\newcommand{\PM}{$\pm$}

\begin{document}

\title{Possible violation of spin-statistics connection in
electron-electron scattering at low relativistic energies} 

\author{R. N. Sen\\Department of Mathematics\\Ben-Gurion
University of the Negev\\Beer Sheva 84105, Israel\\E-mail: {\tt
rsen@cs.bgu.ac.il}}

\twocolumn[

\begin{@twocolumnfalse}
 
\maketitle\thispagestyle{empty}

\begin{abstract}

In 1954, Ashkin, Page and Woodward (hereafter APW) reported on the
first counter experiments to measure the \ee{} and \ep{} scattering
cross-sections at low relativistic energies (0.6--1.7 MeV).  Their aim
was to look for the spin and exchange or virtual annihilation effects
predicted by the \Mol{} and Bhabha formulae. Their experiments
confirmed these effects, but the measured cross-sections at 0.61 MeV
were significantly smaller than their predicted values. The authors
remarked that these deviations were `presumably due to multiple
scattering'.  However, careful reading of the unpublished theses of
Page (1950) and Ashkin (1952), Page's letter (1951) and the APW paper
reveals no credible evidence for multiple scattering at 0.61 MeV; if
anything, the evidence rules \emph{against} multiple scattering.  If
multiple scattering is ruled out, the observations may indicate
\emph{a departure from quantum electrodynamics}.  This departure may
be due to a non-Coulomb central force, a weakening of the
spin-statistics connection, or both.  Only experiment can tell which
of these possibilities holds true, and therefore we suggest that new
\ee{} scattering experiments be carried out at different energies (at
0.4--1.0 MeV) and different scattering angles, as well as specific
tests for multiple scattering.  We consider a non-Coulomb central
force to be very unlikely, and advance the hypothesis that a
fraction of the electron pairs scatter as spin-zero fermions (which
would lower the observed cross-section).  Numerical calculations show
that this hypothesis may be tested quantitatively in a Page-type
experiment, even with little improvement in the accuracy he achieved
in 1950. 

\vspace*{1.8em}

\end{abstract}

\end{@twocolumnfalse}
]

\section{Introduction}\label{INTRO}

The electron-electron (\ee) scattering cross-section first attracted
interest owing to its role in determining the penetrating power of
fast electrons (from cosmic rays and radioactive substances) in their
passage through matter \cite{BIRKHOFF1958}. Shortly after \Mol{}
published his celebrated formula in 1932 \cite{MOLLER1932}, Champion
-- who had been corresponding with \Mol{} -- published the results of
his cloud-chamber experiments on the subject; his finding was that
\Mol{}'s formula fitted the data better than five other candidates
\cite{CHAMPION1932}. Champion did not have sufficient data to make a
stronger assertion; the \Mol{} formula contains terms which arise from
spin and exchange, but their effects are most pronounced at large
scattering angles which are found only in a very small fraction of
scattering events. Therefore in the late 1940s Page and Ashkin,
graduate students of Woodward at Cornell, carried out two separate
experiments at energies between 0.6--1.7 MeV and 0.6--1.2 MeV
respectively, using coincidence counters arranged to detect only
large-angle scattering events.\footnote{Ashkin's chief aim was to
measure the ratio of \ee{} and \ep{} scattering cross-sections, which
could be determined more accurately than either absolute
cross-section.} Details of these experiments were presented in the
Ph~D theses of Page (1950) \cite{PAGE1950} and Ashkin (1952)
\cite{ASHKIN1952}, which remained unpublished, except for a brief
letter by Page in 1951 \cite{PAGE1951}. A summary of their results on
\ee{} and \ep{} scattering was published in 1954 in a joint paper by
Ashkin, Page and Woodward (hereafter APW) \cite{APW1954}. The results
left no doubt that, at large scattering angles, spin and exchange (or
spin and virtual annihilation, for \ep{} scattering)  modified the pure
Coulomb scattering cross-section significantly, as had been suggested
by Oppenheimer \cite{OPPY1928} and Mott \cite{MOTT1930} in the early
years of quantum-mechanical scattering theory.

\vspace*{-1.3cm}
{
\begin{figure}[ht]
\hfil
{
\begin{minipage}[t]{\linewidth}
\hfil\includegraphics*[width=.9\linewidth,keepaspectratio=true]{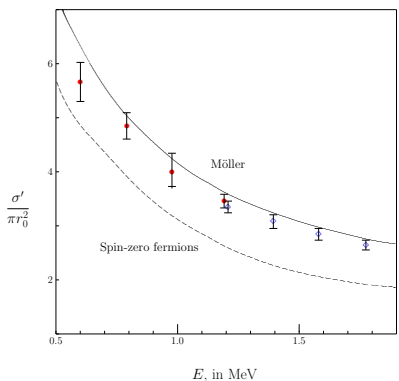}\hfil

\vspace*{-2.5cm}
\caption{$e\mhyphen e$ scattering cross-section at $x=0$
from collodion (\textcolor{red}{$\bullet$}) and beryllium
(\textcolor{blue}{$\diamond$}) foils (after Ashkin, Page and Woodward
\cite{APW1954})}\label{FIG-APW}
\end{minipage}
}
\hfil
\end{figure}
}

\vspace*{-1cm}

The results obtained with the `$270^{\circ}$-apparatus' used by Page
({Fig.}~5 of \cite{APW1954}) are reproduced\footnote{Photographic
reproduction having proven unsatisfactory, we have redrawn the figure
as best as we could.} in {Fig.}~\ref{FIG-APW}. (Ashkin used mylar
foils for his experiments, but his results on \ee-scattering agreed
with those shown in {Fig.}~\ref{FIG-APW}; see Fig.~6 of
\cite{APW1954}.) Notice that the observed value at 0.6 MeV lies
definitely below the \Mol{} curve (solid line).  This discrepancy will
be the centre of our attention because, if it cannot be explained by
experimental errors and/or multiple scattering\footnote{It would be
convenient, for our purposes, to distinguish between multiple
scattering and other sources of error.} then it would imply \emph{a
breakdown of \emph{QED} at this energy}.  APW state that this
discrepancy arose `presumably because of multiple scattering'. (The
same discrepancy was observed at the same energy for \ep{} scattering
by Ashkin; APW use exactly the same phrase to explain it.) In
{Sec.}~\ref{MULTIPLE}, we shall examine the evidence presented by APW
in some detail by going back to the original sources
\cite{PAGE1950,ASHKIN1952,PAGE1951}, and shall conclude that
the empirical data \emph{do not support} the multiple scattering
hypothesis.  Therefore there is a very strong case for \emph{new
experimental studies of the \ee{} cross-section} at energies between
0.4 to 1.0 MeV. The object would be to determine (i)~whether there are
true departures from \Mol's formula, and (ii)~if there are, whether
the observed departures can be explained by the hypothesis that a
fraction of the electron pairs scatter as spin-zero fermions
(discussed in {Sec.}~\ref{ALTERNATIVES}).

The reader may wonder why a gap of such profound theoretical
consequence has not been closed by experiment in the last 65 years.
It would be presumptuous of the present author, who is not a historian
of physics, to try to answer this question.  He can only suggest a few
references that capture some of the excitement, and record some of the
concerns of physicists of the time: (i)~the definitive history of QED
by Schweber \cite{SCHWEBER1994}, (ii)~the book by Bethe and de Hoffman
for the state of `meson physics' in 1954 \cite{BdH1955}, and finally
(iii)~Dyson's account of how he put together his key paper `The
radiation theories of Tomonaga, Schwinger and Feynman'
\cite{DYSON1949} in a Greyhound bus, given towards the end of
chapter~6 of his memoirs \cite{DYSON1979}.  Dyson's paper predated the
publications of Schwinger and Feynman!

The rest of this paper is organised as follows.  In
{Sec.}~\ref{MOLLER} we recall the scattering formulae with which we
shall be concerned. They include the \Mol{} formula, the formulae for
the scattering cross-sections of two spin-zero fermions [sic!]  and
two spin-zero bosons of equal charge, and the Bhabha formula --
suppressing some details -- for \ep{} scattering \cite{BHABHA1936}. We
provide enough material to make the account more or less
self-contained. In {Sec.}~\ref{EXPT} we assemble some data from Page's
thesis and letter (pointing out a significant inconsistency between
the two) and Ashkin's thesis, with emphasis on the multiple scattering
problem. We analyse the data in {Sec.}~\ref{MULTIPLE} and conlude
that, \emph{contrary to the presumption of} APW, the evidence for
multiple scattering at 0.61 MeV, both in \ee{} and \ep{} scattering,
is lacking. We also suggest a direct test for multiple scattering.  In
{Sec.}~\ref{ALTERNATIVES} we try to devise an alternative explanation,
and advance the hypothesis that with decreasing energy, an increasing
fraction of the electron pairs scatter as spin-zero fermions
(Hypothesis I, eq.\ (\ref{EQ-HYPO})).  In {Sec.}~\ref{ESTIMATES} we
make numerical estimates of the experimental accuracy that would be
required to put this hypothesis to test in off-foil scattering
experiments, and conclude that it is indeed possible to do so even
with little improvement on Page's experiment of the late 1940s. (A
further test for multiple scattering emerges in the process.)  In
section \ref{FREE-FREE}, we give some details about the colliding beam
experiment of Williams et al (published 2014); an experiment with
suitable beams may greatly reduce the possibility of errors due to
multiple scattering, but we are unable to assess its feasibility at
low relativistic energies. In the next section we comment on the
far-reaching theoretical implications of hypothesis (\ref{EQ-HYPO})
being validated by experiment.  In the last section we provide a brief
sketch of a notion of `small' violations of the Pauli principle in
bound states and of the ongoing experiments to detect such violations,
with references for the interested reader. An appendix gives the
numerical tables on which the estimates of Sec.\ \ref{ESTIMATES} are
based.


\section{Scattering formulae}\label{MOLLER}

When a particle of positive mass is scattered by another at
relativistic energies, the differential cross-section for the process
is most easily calculated in the centre-of-momentum frame. One would
expect that it has then to be transformed to the laboratory frame, in
which the target particle is at rest, for use by the experimentalist.
This transformation can be quite complicated in relativistic
kinematics, even when both particles have the same mass. But \Mol{}
made the felicitous observation that it was not necessary to carry out
this transformation; the scattering angle in the centre-of-momentum
frame was very simply related to the fractional energy transfer --
which could be observed directly -- from the incident to the target
particle, and he communicated his results by letter to Champion, who
was performing the experiment (see \cite{ROQUE1992}). 

Champion wrote down \Mol{}'s formula as follows (see
\cite{CHAMPION1932}, eq.\ 2); he noted with apparent surprise
that Planck's constant appears nowhere in it!

\begin{equation}\label{EQ-MOLLER}
\begin{array}{rcl}
\sigma^{\prime}_{\sf M}&=&\dfrac{\rmd\sigma_{\sf M}(x)}{\rmd x}\\[1em] 
&=& 4\pi \left(\dfrac{r_0}{\beta^2}\right)^2 
\dfrac{\gamma+1}{\gamma^2}\left[\dfrac{4}{(1-x^2)^2} -
\dfrac{3}{1-x^2}\right.\\[1.5em]
&&+\left.\dfrac{(\gamma-1)^2}{4\gamma^2}\left(1+\dfrac{4}{1-x^2}\right)\right]
\end{array}
\end{equation}

\vspace{1em}\noindent
In the above, the independent variable $x$ is the cosine of the
scattering angle ($x=\cos\theta_{\sf cm}$) in the
\emph{centre-of-momentum system}. The quantity $\sigma^{\prime}_{\sf
M}$ ({\sf M} for \Mol{}) is formally the $x$-derivative of the total
cross-section $\sigma_{\sf M}$ in the laboratory system,
$r_0=e^2/mc^2$ is the classical radius of the electron (in CGS units),
$v$ is the velocity of the incident electron, $\beta=v/c$ and
$\gamma=(1-\beta^2)^{-1/2}$.  Finally, $x$ is related to $\theta$, the
scattering angle in the laboratory system, by the complicated
expression
\begin{equation}\label{IND-VAR}
 x = \cos\theta_{\sf cm} =\dfrac{2-(\gamma+3)\sin^2\theta}
{2+(\gamma+1)\sin^2\theta}{}
\end{equation}

\noindent where $\theta_{\sf cm}$ is the scattering angle in the
centre-of-momentum system. What makes (\ref{EQ-MOLLER}) usable by the
experimentalist is \Mol{}'s finding, mentioned above: the variable $x$
may be expressed quite simply in terms of $w$, the fraction of kinetic
energy of the incident electron transferred to the target electron
(assumed at rest):\footnote{{See \Mol{}'s original article
\cite{MOLLER1932}, page 569, the second unnumbered equation between
the two equations each numbered (75$^{\prime}$).}}
\begin{equation}\label{X}
x = 1 - 2w
\end{equation} 

If one calculates the same cross-section for a pair of spin-zero
fermions \cite{MM1965} -- a physical impossibility if the
spin-statistics theorem holds -- one finds the same formula, but
without the term
\begin{equation}\label{SPIN}
\dfrac{(\gamma-1)^2}{4\gamma^2}\left(1+\dfrac{4}{1-x^2}\right)
\end{equation}
inside the square brackets in (\ref{EQ-MOLLER}). This term ``may thus
be considered to be the contribution made by the spin'' of the
electron (\cite{MM1965}, page 817).  If the last two terms in the
square brackets in (\ref{EQ-MOLLER}) are dropped, what remains may be
called the relativistic Rutherford formula with $Z=Z^{\prime}=1$ and
$m=m^{\prime}$; we shall denote this cross-section by $\sigma_{\sf
R}^{\prime}$.  Likewise, we shall denote the cross-section for the
scattering of two charged spin-zero fermions by $\sigma_{\sf
F}^{\prime}$.

Spin-zero bosons ({\sf szb}) exist in nature. The electrodynamics of
spin-zero particles (scalar electrodynamics) was constructed in 1950
(see \cite{MATTHEWS1950,ROHRLICH1950}). The lowest order scattering
cross-section for two such particles of equal charge -- in the
laboratory system -- turns out to differ from the \Mol{} formula
(\ref{MOLLER}) only in the last two terms contained in the square
brackets (see \cite{IZ1985}, page 286). The full formula is

\begin{equation}\label{SCALAR-B}
\begin{array}{l}
\sigma^{\prime}_{\sf szb} = \dfrac{\rmd\sigma_{\sf szb}(x)}{\rmd x} =
\\[1.5em] 
\quad\quad 4\pi \left(\dfrac{r_0}{\beta^2}\right)^2 
\dfrac{\gamma+1}{\gamma^2}\left[\dfrac{2}{1-x^2} -
\dfrac{\gamma-1}{2\gamma}\right]^2
\end{array}
\end{equation}

In the standard textbook and reference work by Jauch and Rohrlich
\cite{JR1955}, the \Mol{} formula is written slightly differently.
They use the natural system of units in which $c=\hbar=1$ and length
is measured in centimeters. They also measure the electric charge in
the rationalized system of units, in which $e^2/(4\pi c\hbar) \approx
1/137$.) Their version is

\begin{equation}\label{MOLLER-JR}
\begin{array}{rcl}
\dfrac{\rmd\sigma_{\sf M}}{\rmd\Omega} &=&
r_0^2\left({4}\dfrac{\gamma+1}{\beta^2\gamma}\right)^2
     \dfrac{\cos\,\theta}{[2+(\gamma-1)\sin^2\theta]} \\[1em]
&&\times\quad
{\big[\sf \Mol\big]}
\end{array}
\end{equation}
where the quantity [{\sf \Mol}] is \emph{exactly} the same as the
quantity inside the large square brackets in (\ref{EQ-MOLLER}).
Neither Champion nor APW have made use of the formula in this form,
but for us the relevant fact is that Jauch and Rohrlich have also
derived Bhabha's formula for \ep{} scattering cross-section using the
same notation. Denoting this cross-section (again, in the laboratory
system) by $\sigma_{\sf B}$, the formula given by Jauch and Rohrlich
may be written as follows:

\begin{equation}\label{BHABHA-JR}
\begin{array}{rcl}
\dfrac{\rmd\sigma_{\sf B}}{\rmd\Omega} &=&
r_0^2\left({2}\dfrac{\gamma+1}{\beta^2\gamma}\right)^2
     \dfrac{\cos\,\theta}{[2+(\gamma-1)\sin^2\theta]} \\[1em]
&&\times\quad
\big[{\sf Bhabha}\big]
\end{array}
\end{equation}

\noindent In the above, [{\sf Bhabha}] consists of three separate
terms which are functions of $x$ rather than $x^2$, as the two
particles are not identical. We shall not write them down explicitly;
the full formula may be found on p.\ 260 of \cite{JR1955}. Bhabha
identified the first of these with the scattering of two
spin-$\frac12$ particles of opposite charge that are not antiparticles
of each other.  The second is the virtual annihilation term, and the
third arises from interference between the two. All three are
functions of $x$ and $\gamma$ (or $E$) only, and involve no physical
constants. The ratio

\begin{equation}\label{RATIO-M-B}
\begin{array}{rcl}
\rho(E,x)&=&\dfrac{\rmd\sigma_{\sf M}}{\rmd\Omega}\!\left/
\dfrac{\rmd\sigma_{\sf B}} {\rmd\Omega}\right. =
\dfrac{\sigma^{\prime}_{\sf M}(x)} {\sigma^{\prime}_{\sf
B}(x)}\\[5mm] &=& 4\dfrac{\text{[{\sf \Mol}]}} {\text{[{\sf
Bhabha}]}}
\end{array}
\end{equation} 
remains well-defined as $x\rightarrow 1$, unlike the individual
cross-sections.


\subsection{Nonrelativistic limits}\label{NR-LIMITS}

All formulae for scattering cross-sections given above contain $r_0$
and $\beta$ in the combination $r_0/\beta^2 = e^2/mv^2$,  independent of
$c$. Therefore calculation of the nonrelativistic limit
$c\rightarrow\infty$ is reduced to the substitution $\gamma=1$.  In
this limit the \Mol{} formula~(\ref{EQ-MOLLER}) loses the last term in
the square brackets, and \emph{the resulting formula is identical with
the $($NR limit of\,$)$ the scattering formula for two spin-zero
fermions} derived by Mott.  Similarly, formula (\ref{SCALAR-B}) for
the scattering of two spin-zero bosons (of the same mass and charge)
reduces to the Rutherford formula.  At nonrelativistic energies,
(i)~an electron-electron scattering experiment cannot distinguish
between spin-zero and spin-half fermions, and (ii)~a scattering
experiment cannot distinguish between two scalar bosons and two
classical particles.  The unwritten term [{\sf Bhabha}] in
(\ref{BHABHA-JR}) reduces, likewise, to 
$$ \dfrac{4}{(1-x)^2}$$
The nonrelativistic limit of the ratio (\ref{RATIO-M-B}) reduces to

\begin{equation}\label{NR-LIM-RATIO}
 \dfrac{\sigma^{\prime}_{\sf M}(x)}
{\sigma^{\prime}_{\sf B}(x)}\bigg|_{\sf NR} = \dfrac{1+3x^2}{(1+x)^2}
\end{equation}
%


\section{\mbox{The experiments of Page} and Ashkin}\label{EXPT} 

In their paper, APW give a general overview of the experiments of Page
and Ashkin. At the end of page 360 of \cite{APW1954}, they say: `Care
was taken to avoid errors due to multiple scattering in the scattering
foil', and describe tests to detect its presence. In this section we
shall provide further details taken from the unpublished theses of
Page \cite{PAGE1950} and Ashkin \cite{ASHKIN1952}. We shall also refer
to the figure in Page's brief communication \cite{PAGE1951}. The aim
is to prepare the ground for a more detailed discussion, in
{Sec.}~\ref{MULTIPLE}, of the effect of multiple scattering on the
observed cross-sections.


\subsection{Page's thesis and letter}\label{PAGE}

Page's letter \cite{PAGE1951} contradicts his thesis in a very
significant manner. We shall discuss this after presenting some of his
conclusions on the multiple scattering problem.

\subsubsection{\mbox{Page on the multiple scattering} problem}

Page measured the \ee{} scattering cross-sections at 0.6, 0.8, 1.0 and
1.2 MeV with both beryllium and collodion foils, of densities
4.5\,mg/cm$^2$ and 0.5\,mg/cm$^2$ respectively. Since nothing except
the foil was changed during the experiments, the ratio of the count
rates $C_{\sf Be}/C_{\sf coll}$ at any energy should have depended --
had there been no multiple scattering -- only on the ratio of the
densities of scatterers in the two foils, which was a constant.
Page's plot of $C_{\sf Be}/C_{\sf coll}$ against energy (his Fig.\
$4^\prime$) is shown as the upper graph in our {Fig.}~\ref{FIG-PAGE}.
Had there been no multiple scattering, these points would have lain on
the horizontal line marked in the figure.  Page's comment on it is as
follows ({p.}~41 of his thesis, emphasis added):

\begin{quote}
`In the absence of\ldots{}precise formulas for multiple scattering
(\emph{and clearly any side experiment aiming at such determination
could easily dwarf the main experiment here})\ldots one should simply
accept the observed coincidence rate ratios ({Fig.}~$4^{\prime}$)
between thick and thin foils as the guide whereby certain of the data
is rejected on nuclear scattering grounds.'
\end{quote}
The evidence presented in Page's Fig.~$4^{\prime}$ clearly shows that,
relative to the collodion foil, multiple scattering effects become
progressively more important with decreasing energy in the beryllium
foil.  Use the word `nuclear' rather than `multiple' in the last
sentence may require justification, but the point is not relevant to
our discussion, because Page's final data at the lower energies (0.6,
0.8 and 1.0 MeV) were obtained using the collodion foil.

{
\begin{figure}[ht]
\hfil
{
\begin{minipage}[t]{\linewidth}
\hfil\includegraphics*[width=.9\linewidth,keepaspectratio=true]{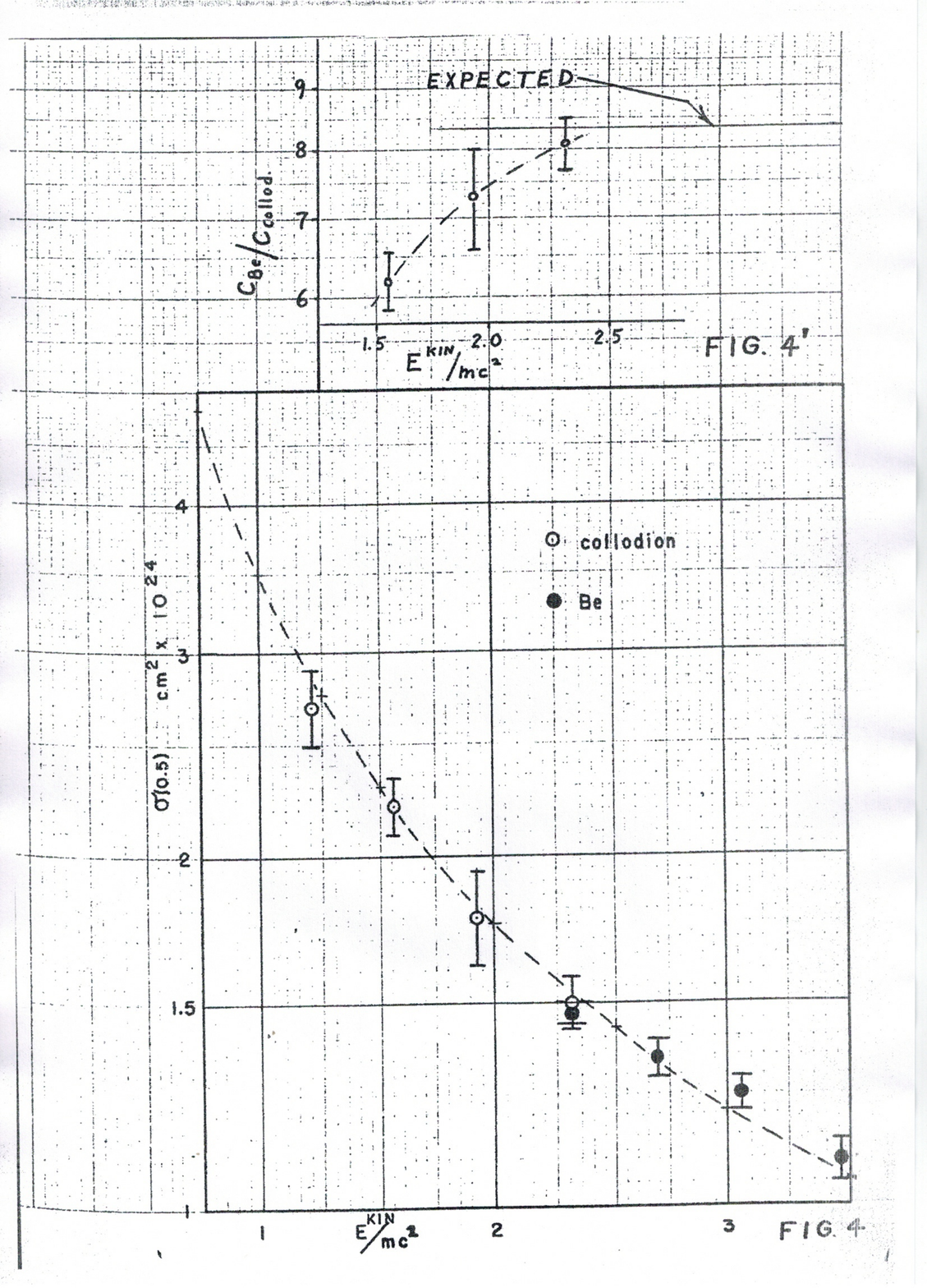}\hfil
\caption{Page's Fig.\ 4 (lower figure) and Fig.\ 4$^{\prime}$ (upper
figure)}\label{FIG-PAGE}
\end{minipage}
}
\hfil
\end{figure}
}

Page's Fig.\ 4 and Fig.\ $4^{\prime}$ are reproduced in our
Fig.~\ref{FIG-PAGE}. (Note that Page expresses $E$ as multiples of
$mc^2$, i.e., in units of 0.511 MeV.) All the observed points on his
his {Fig.}~4 are close to the \Mol{} curve. On page~42 of his thesis,
he says:

\begin{quote}
`\ldots which implies that all the data of {Fig.}~4 is for all
practical purposes free of multiple scattering loss.'
\end{quote}

\noindent However, the last assertion was contradicted (implicitly) by
his letter, and is discussed below.

\subsubsection{The inconsistency}

The raw data obtained by Page were used to draw three different energy
vs cross-section graphs: (i)~{Fig.}~4 in Page's thesis (lower graph of
our Fig.\ \ref{FIG-PAGE}), (ii)~{Fig.}~2 in Page's letter
\cite{PAGE1951} (which we have not reproduced) and (iii)~APW's
{Fig.}~5 (our {Fig.}~\ref{FIG-APW}).  Even a quick glance shows that
(i) and (iii) do not agree. In (i), the data points seem to be
\emph{shifted upwards} with respect to those in (iii), and the points
at the two highest energies lie \emph{clearly above} the \Mol{} curve.
The graphs in (ii) and (iii) do agree with each other.  The remark
from p.\ 42 of Page's thesis quoted above is not applicable to the
graphs in (ii) and (iii), and is inconsistent with statements in APW.
This suggests that at least one of the two formulae used for
converting count rates to cross-sections given on pages 56 and 57 of
Page's thesis was slightly modified to arrive at the graphs in (ii)
and (iii).  But no explanation is offered either in Page's letter
\cite{PAGE1951} or in APW; there is no reference to this inconsistency
in either publication.

In view of this unexplained inconsistency, we shall use the graphs in
(ii) and (iii) (our {Fig.}~\ref{FIG-APW}) as describing the results of
Page's experiments, rather than the one presented in his thesis,
except when stated otherwise.


\subsection{Ashkin's thesis}\label{ASHKIN}

Ashkin's apparatus was similar to Page's, but somewhat larger, to
allow for lead shielding. (His positron source, Co$^{56}$, emitted
three $\gamma$ rays for every positron.) Exigencies of shielding also
restricted the trajectories of the scattered particles to
$180^{\circ}$, as opposed to Page's $270^{\circ}$. For \ep{}
scattering, both counters were on the same side of the partition; for
\ee{} scattering, the magnetic field had to be reversed and one of the
counters moved to the other side of the partition. (See {Fig.}~1 of
Ashkin's thesis or {Fig.}~3 of APW.) Thus the \ee{} scattering
experiments of Page and Ashkin could be considered `essentially
different'. Note that Ashkin used Mylar foils of densities
1.7\,mg/cm$^2$ and 0.9\,mg/cm$^2$ as scatterers (p.~14 of
\cite{ASHKIN1952}).

Ashkin measured the \ee{} scattering cross-section at $E=1.220,\;
1.019,\; 0.818\;\text{and}\; 0.611$ MeV and $x=0$. At the two higher
energies, he used only the thicker foil, at 0.611 MeV, only the
thinner foil, and both foils at 0.818 MeV. At this energy, the
cross-section obtained with the thinner foil was 4.7\% larger than
that obtained with the thicker one, which he took as evidence of
multiple scattering in the thicker foil.  At 0.611 MeV, the value he
obtained $(2.60\times10^{-24}\,\mathrm{cm}^2)$ was about 3\% smaller
than the one obtained by Page $(2.69\times10^{-24}\,\mathrm{cm}^2)$
with the 0.5\,mg/cm$^2$ collodion foil and presented in his thesis
(see Table II, p.~14 of \cite{ASHKIN1952} and Table III, p.~14 of
\cite{PAGE1950}). 

But, as we have noted earlier, the calculated cross-sections given in
Page's thesis and used in his {Fig.}~4 (our {Fig.}~2) do not agree
with {Fig.}~5 of APW (our {Fig.}~1).  Ashkin's results are shown in
graphical form in {Fig.}~6 in APW. This can be compared visually
with {Fig.}~5 of APW, which is based on Page's data. One sees that, at
0.6 MeV, the cross-section found by Ashkin is, if anything,
\emph{slightly larger} than the one shown in {Fig.}~5 of APW. As APW
state in the penultimate paragraph on page 360, the results of Ashkin
`agreed well' with those of Page.

Ashkin also determined the ratio $\rho(E,x)$ of \ee{} and \ep{}
cross-sections (\ref{RATIO-M-B}) at $x=0$ and $E=1.02,\; 0.82\;
\text{and}\; 0.61$ MeV directly from the coincidence counts, and
compared the observed values with those calculated from \Mol's and
Bhabha's formulae. (Indeed, this was the main aim of his thesis.) He
observed that most of the experimental errors, due to limitations of
the apparatus, would cancel each other in the ratio. (We shall exploit
this observation in {Sec.}~\ref{ESTIMATES}.)  The agreement between
theory and experiment was indeed remarkable.  (We shall present his
data in Table \ref{ASHKIN-TABLE}.)  After presenting his data (p.~16
of \cite{ASHKIN1952}), he wrote:

\begin{quote}
`Even the 0.61 MeV point lies right on the theoretical curve. \emph{This
serves to substantiate the supposition that both absolute numbers for
the $[$\mbox{\ee} and \ep$]$ cross-sections at this energy were low due to
multiple scattering}' [our emphasis].
\end{quote}

\noindent The same claim was repeated in APW. Neither Ashkin nor APW
offered any justification for it. We shall examine it in
{Sec.}~\ref{EQUALITY}. 


\section{\mbox{The multiple scattering} problem}\label{MULTIPLE}

The experiments of Page and Ashkin were based on the implicit
assumption that \emph{single scattering dominates over other
processes}.  Their results amply justify this assumption.  {Fig.\
\ref{FIG-APW}} also suggests that the ratio of the observed \ee{}
cross-section to the calculated \Mol{} cross-section \emph{decreases}
with decreasing energy below 1.0 MeV.  APW and Ashkin suggest that
this effect (which turned out to be of little relevance to their main
aim of verifying the existence of the exchange/virtual annihilation
terms in the cross-sections for \Mol/Bhabha scattering respectively)
may be due to multiple scattering. The evidence they offered was of
two kinds: (i)~the effect of foil thickness, and (ii)~the equality of
observed and calculated ratios of the \ee{} and \ep{} cross-sections
at certain energies.  We shall consider them separately.


\subsection{Effect of foil thickness }

We shall now bring together the relevant data from the original
sources, and set down the conclusions that may be drawn from them. 

\begin{enumerate}

\item At 0.82 MeV and $x=0$, the cross-section measured with a
1.7\,mg/cm$^2$ Mylar foil is 4.7\% smaller than that measured with a
0.9\,mg/cm$^2$ Mylar foil (Ashkin \cite{ASHKIN1952}). This indicates
that the influence of multiple scattering increases with increasing
thickness of foils of the same material.

\item At 0.61 MeV and $x=0$, the cross-section measured with a
0.9\,mg/cm$^2$ Mylar foil by Ashkin \cite{ASHKIN1952} is 3\% smaller
than that measured with a 0.5\,mg/cm$^2$ collodion foil by Page (as
reported in his thesis \cite{PAGE1950}, Table III, page 14). However,
as pointed out earlier, the data of Page's Table III are inconsistent
with {Fig.}~5 of APW, and a visual comparison of {Figs.}~5~and~6 of
APW suggests that the cross-section obtained by Ashkin with the
0.9\,mg/cm$^2$ Mylar foil is, if anything, \emph{slightly larger} than
that obtained by Page with the 0.5\,mg/cm$^2$ collodion foil.

\end{enumerate} 

\noindent Item 1 of the above is supportive of the idea that
multiple scattering reduces the observed cross-section, at least in
the thicker foil. Item 2 is more problematic; if we accept that APW
overrides Page's thesis, the near-equality of the cross-section
measured with two different foils, one of which is nearly twice as
dense as the other, would seem to indicate the \emph{absence} of
multiple scattering in both the experiments. Then \emph{the low value
of the cross-section \emph{(at 0.61 MeV)} remains to be explained}. It
should be emphasized that, in both cases, the data are too meager to
draw hard conclusions.


\subsubsection{Test for multiple scattering, I}\label{TMS-I}

The experiment, which can be a variant of Page's experiment, would be
carried out at fixed $E \approx{} 0.5$--$0.6\,\text{MeV}$ and $x=0$,
but with foils of different thicknesses made out of the same material,
the thinnest having a lower density than Page's collodion foil. All
other parts of the apparatus will remain unchanged. The idea would be
to obtain enough points in a plot of the cross section against (the
decreasing) foil density to detect whether or not the cross section
tends to level off at lower foil thicknesses.  The hypothesis of
dominance of single scattering would suggest that the cross section
should level off with decreasing foil density.  If it levels off at
the \Mol{} value, it would be strong evidence that multiple scattering
alone suffices to explain the phenomenon. On the other hand, if it
levels off at a value \emph{smaller} than the \Mol{} value, it would
be strongly suggestive of a different mechanism, such as the one to be
considered in Sec.~\ref{ALTERNATIVES}, being active either
alternatively or simultaneously. 

The feasibility of the experiment would depend on the fabricability of
suitable scattering foils. The range of thicknesses should be such
that the levelling-off effect is observable. \emph{If such foils can
be made, the experiment would be a decisive one.}


\subsection{Equality of observed and calculated ratios of $\spr_{\sf
M}/\spr_{\sf B}$}\label{EQUALITY}

Recall that Ashkin's main aim was to determine the ratios $\rho(E,x)$
of the \Mol{} and Bhabha cross-sections directly from the count ratios
(using the same apparatus); many of the errors in the determination of
the absolute cross-sections would cancel out in the ratio. If
\emph{all} errors cancelled out, the observed ratio would equal the
calculated ratio. As pointed out in {Sec.}~\ref{ASHKIN}, the agreement
was near-perfect at $E$ = 1.02 and 0.61 MeV (at $x=0$), and very good
at $E$ = 0.82 MeV. We shall analyze Ashkin's `supposition' that this
near-equality explains the low observed values for the \ee{} and \ep{}
cross-sections at $E=0.61$\;MeV. Our analysis will be based on the
observation that the `dominance of single scattering' implies that
corrections to single scattering will, in turn, be dominated by a
single, second collision. 

The cross-sections $\sigma^{\prime}_{\sf M}(E,x)$ and
$\sigma^{\prime}_{\sf B}(E,x)$ diverge at $\theta=0\, (x=1)$ and then
decrease very rapidly with increasing $\theta$ (decreasing $x$).  As a
result, a particle that suffers a second collision in the foil will
most probably be deviated only very slightly from its path, which may
not be enough to prevent it from striking the counter. Only particles
that suffer larger, less-probable deviations will avoid striking the
counter. Owing to the divergence of the cross-section, the fraction of
particles (electrons or positrons) that are prevented from reaching
the counter due to a second collision in the foil cannot be estimated.
However, \emph{the ratio} $\rho(E,x)$ of the \ee{} and \ep{}
cross-sections defined by (\ref{RATIO-M-B}) remains finite as
$x\rightarrow0$, and we can make a rough estimate of the effect of a
second collision using this ratio -- and the geometry of the apparatus
-- as follows.

We assume that an incoming electron or positron of energy $E$ is
scattered by an electron in the foil (considered at rest), imparting
half its kinetic energy to the latter -- i.e., $x=0$ for the scattered
particles. Next, \emph{one} of the scattered particles suffers a
second collision in the foil which changes its direction of flight by
a small angle $\ths$ which does not materially affect its
time-of-flight to the counter.  (The variables referring to the second
collision will be distinguished by the subscript `{\sf s}'.) \emph{If
all second collisions scatter by the angle $\theta_{\sf s}$}, their
effect will be to multiply the \Mol{} cross-section $\sigma_{\sf
M}^{\prime}(E, x=0)$ by a factor proportional to $\sigma_{\sf
M}^{\prime}(\Es,\xs)$, where $\Es = E/2$, $\xs=1-\delta$ and $\delta$,
only slightly greater than zero, is determined by the geometry of the
apparatus.  For positron scattering, the Bhabha cross-section
$\sigma_{\sf B}^{\prime}(E, x=0)$ will be modified by a factor
proportional to $\sigma_{\sf B}^{\prime} (\Es,\xs)$. If, as in
Ashkin's experiments, the same apparatus is used for both experiments,
the two proportionality factors should be the same.  We may therefore
\emph{assume} that the ratio $\rho(E,x=0)$ will be multiplied by the
factor

\begin{equation}\label{DEFINE-F}
F(\Es,\xs) = \dfrac{\sigma_{\sf M}^{\prime}(\Es,
\xs)}{\sigma_{\sf B}^{\prime}(\Es,\xs)}
\end{equation}
where $\Es=E/2$ and $\xs=1-\delta$. 

\begin{figure}[ht]
\hfil
{
\begin{minipage}[t]{\linewidth}
\vspace{2ex}
\hfil\includegraphics*[width=0.9\linewidth,keepaspectratio=true]{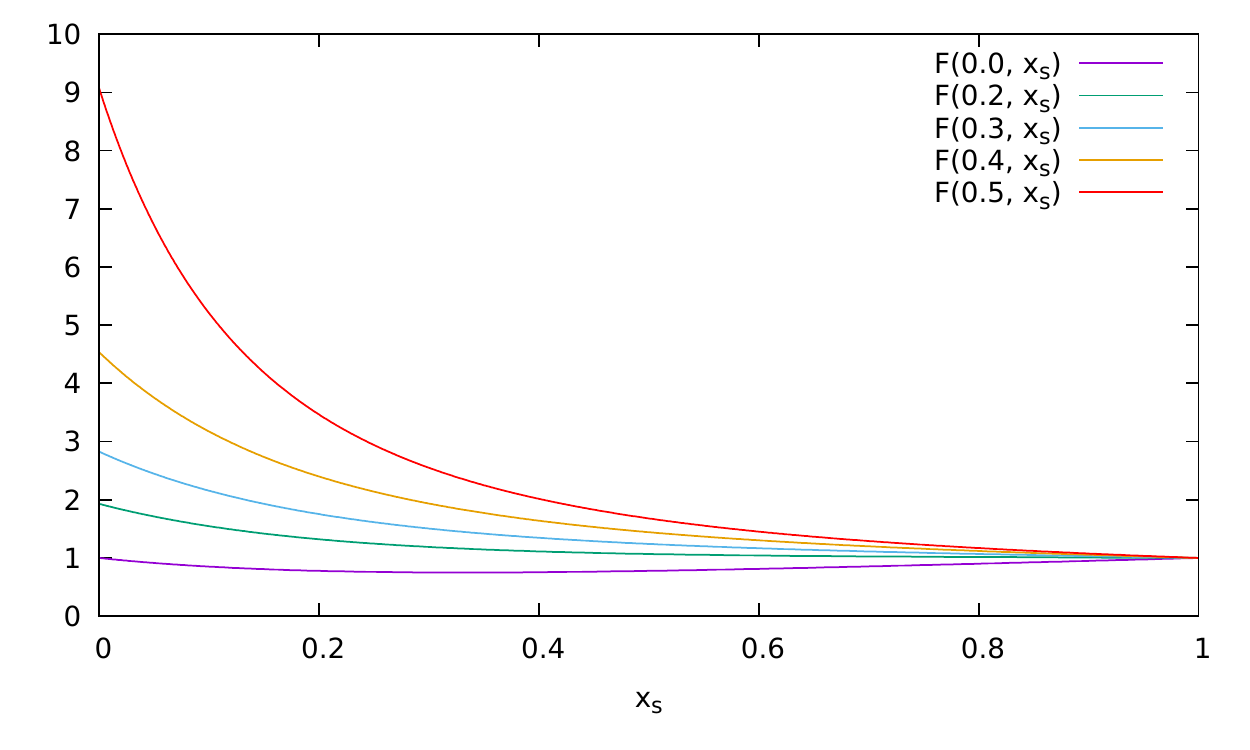}\hfil
\caption{Graphs of the ratio $F(\Es,\xs)$ vs $\xs$ for $\xs\in [0,1]$
and $\Es$ = 0.0, 0.2, 0.3, 0.4 and 0.5 MeV [in grey scale, bottom to
top on the $x_{\sf s}=0$ axis].}\label{FIG-LARGE}
\end{minipage}
}
\hfil
\end{figure}

\begin{figure}[ht]
\hfil
{
\begin{minipage}[t]{\linewidth}
\vspace{2ex}
\hfil\includegraphics*[width=0.9\linewidth,keepaspectratio=true]{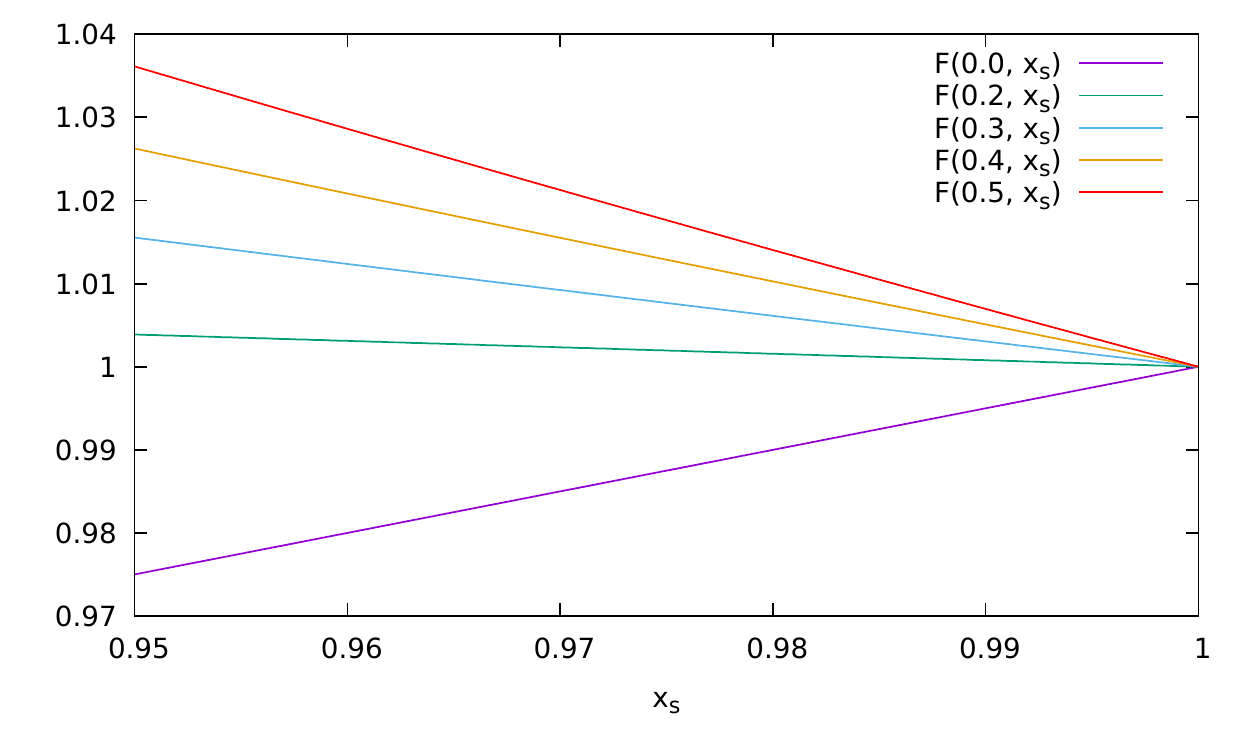}\hfil
\caption{Graphs of the ratio $F(\Es,\xs)$ vs $\xs$ for $\xs\in [0.95,1]$ and
$\Es$ = 0.0, 0.2, 0.3, 0.4 and 0.5 MeV [in grey scale, bottom to top
on the $x_{\sf s}=0.95$ axis].}\label{FIG-SMALL}
\end{minipage}
}
\hfil
\end{figure}

Fig.\ \ref{FIG-LARGE} shows the graph of the ratio $F(\Es,\xs)$ for
$\xs\in [0,1]$ and $\Es = 0.0,\, 0.2,\, 0.3,\, 0.4\;\text{and}\;0.5$
MeV. Fig.\ \ref{FIG-SMALL} is a magnification of the above in the
range $xs\in[0.95,1]$. Except at $\xs=0$, where they all meet, the
graphs are totally disjoint, with those for higher $E$ lying entirely
above those for lower $E$. The case $E=0$ corresponds to the
nonrelativistic limit, and its graph is that of the function
(\ref{NR-LIM-RATIO}).  Except for $E=0$, the graphs decrease
monotonically from $\xs=0$ to $\xs=1$.

The scattering angle $\ths$ (in the laboratory system) can be
calculated for given $\xs$ from (\ref{IND-VAR}).  Table
\ref{TABLE-ANGLE} shows the values of $\ths$, in degrees, calculated
using (\ref{IND-VAR}) for $\xs$=0.995, 0.99, 0.98, 0.97 and 0.95 and
$\Es = 0.2,\,0.3,\,0.4\; \text{and}\; 0.5\;\text{MeV}$.  Recall that
$\Es=E/2$, where $E$ is the energy of the electron or positron
incident upon the foil.

\newcolumntype{d}[0]{D{.}{.}{2}}
\begin{table}[h]

\small
\begin{center}
\begin{tabular}{|c|d|d|d|d|}\hline
    &\multicolumn{4}{c|}{$\Es$, in MeV}\\ \cline{2-5}
\raisebox{1em}{${\xs}$}   & 0.2  & 0.3  & 0.4  & 0.5\\[1em]\hline
0.995& 2.62 & 2.52 & 2.43 & 2.35 \\%
0.990& 3.71 & 3.57 & 3.44 & 3.32\\%
0.980& 5.25 & 5.05 & 4.87 & 4.71\\ 
0.970& 6.44 & 6.19 & 5.97 & 5.55\\%
0.950& 8.33 & 8.01 & 7.71 & 7.48 \\ \hline
\end{tabular}
\caption{The scattering angle $\ths$ (in degrees) as a function of
$\Es$, for $\xs$ = 0.995, 0.990, 0.980, 0.970 {and} 0.950}
\label{TABLE-ANGLE}
\end{center}
\end{table}

Finally, let us consider the geometry of the apparatus.  The
counters used by Ashkin were made of 0.8-inch (about 2 cm) square
tubing (\cite{APW1954}, p.\ 359). A very rough estimate based on
{Fig.}~1 of Ashkin's thesis shows that, for $x=0$, the more energetic
particles follow a trajectory about 30--40\,cm long before reaching
the counters. In the plane, a circular arc of length 2\,cm subtends an
angle of about 4$^{\circ}$ at 30\,cm, and we shall not be far off in
assuming that this remains true even for the helical paths that the
particles traverse in the experiment. That is, the angular aperture of
the detector at the point of scattering is about 4$^{\circ}$. (For
less energetic particles, the angular aperture will be smaller if the
same counters are used.)  Particles that are scattered \emph{into}
this angular aperture are counted; those that are scattered \emph{out
of it} are not.

We shall now try to put together the assumptions made and
 information obtained so far. The key assumption is that the
effect of multiple scattering can be well approximated by the more
precisely quantifiable idea of double scattering. The effect of a
second collision that changes the direction of flight of a particle by
an amount determined by $\xs$ is to multiply the ratio $\rho(E,x)$
defined by (\ref{RATIO-M-B}) by the factor $F(\Es,\xs)$ defined by
(\ref{DEFINE-F}). To take all possible directions into account in the
second scattering, we have to average $F(\Es,\xs)$ over the interval
$[0,\xs]$. This average is simply the area under the curve devided by
the length of the interval. From Table~\ref{TABLE-ANGLE} and the
angular aperture of the counter estimated earlier, we see that for
$x_s$ around 0.99 or less, almost all the particles being scattered a
second time would be reaching the counter. We now see from Figs.\
\ref{FIG-LARGE} and \ref{FIG-SMALL} that the contributions of the even
larger interval $\xs\in[0.98,1]$ to the areas under the curves for
$\Es \geq 0.2$~MeV is entirely negligible. Therefore, for each $\Es$,
the weighting factor becomes the area under the curve in the interval
$[0,1]$. These areas have to be computed numerically for each $\Es$.

However, we do not have to carry out these numerical integrations. It
is clear from Figs.\ \ref{FIG-LARGE} and \ref{FIG-SMALL} that
$F(\Es,\xs)$ is always greater than unity. Therefore the area under
the curve will be significantly greater than unity, and \emph{will
increase significantly with increasing $\Es$}.

\begin{table}[h]

\small
\begin{center}
\begin{tabular}{|c|c|c|c|}\hline
    &\multicolumn{3}{c|}{$\rho$, observed and calculated}\\
    \cline{2-4}
{$E$, MeV}   & $\rho_{\sf obs}$  & $\rho_{\sf calc}$  & $\rho_{\sf
obs}/\rho_{\sf calc}$\\[2mm]\hline

1.02& $3.74\pm0.49$ & 3.76 & 1.0053 \\
0.82& $3.29\pm0.19$ & 3.42 & 1.0395 \\ 
0.61& $3.09\pm0.22$ & 3.08 & 0.9968 \\ \hline
\end{tabular}
\caption{Observed and calculated values of $\rho$ (Ashkin) and their
ratio}
\label{ASHKIN-TABLE}
\end{center}
\end{table}

Table \ref{ASHKIN-TABLE} gives the ratios $\rho(E,x=0)$ as observed
and calculated by Ashkin, as well as the ratios $\rho_{\sf obs}/
\rho_{\sf calc}$. According to our estimates, this ratio, being the
area under the curve of $F{(\Es,\xs)}$, should be (i)~significantly
greater than unity, and (ii)~an increasing function of $\Es$. Ashkin's
data meet neither of these conditions. While three data points are not
enough to reach a definitive conclusion, the data provided
\emph{certainly do not suggest} that the low values of the observed
\ee{} and \ep{} cross-sections are the results of multiple scattering.
They may even suggest the contrary: that the near-equality of the
observed and calculated ratios at $E=0.61$ and 1.02 MeV are indicative
of the insignificance of multiple scattering! But this putative
conclusion is challenged by the data at $E=0.82$ MeV.


\subsection{Summary of this Section}\label{SUMMARY}

We may sum up the discussion in this Section as follows: 

\begin{enumerate}

\item Variation of the cross-sections with foil thickness -- the foils
being of the same material -- does suggest the possibility of multiple
scattering, while not ruling out other explanations. Experiments with
foils of different thicknesses but the same material may be able to
discriminate between multiple scattering and \emph{other possible
causes}.

\item The cross-section at 0.61 MeV, measured with a mylar foil
of density 0.9\,mg/cm$^2$, is about equal or slightly larger than the
one measured with collodion foil of density 0.5\,mg/cm$^2$. This
suggests that multiple scattering is no longer effective at these
densities. Then the fact that the measured cross-section is lower than
its \Mol{} value remains unexplained. The data are too meager to draw
firm conclusions.

\item The data presented by Ashkin on the comparison of observed and
calculated values of $\spr_{\sf M}/\spr_{\sf B}$ are not sufficient to
establish a reason for the observed low values of $\spr_{\sf M}$ and
$\spr_{\sf B}$ at $E=0.61$ MeV and $x=0$. The near-equality of the
observed and calculated ratios at 0.61 and 1.02 MeV may even be taken
to mean that multiple scattering is \emph{not significant} at either
energy, but the data are too meager to draw firm conclusions.

\end{enumerate}
 
We need more experimental data to draw any conclusion.


\section{The search for alternatives}\label{ALTERNATIVES}

If further experiments show that the low-energy large-angle scattering
cross-sections are unquestionably smaller than their \Mol{} values,
what could be the possible explanations? 

The electron-electron interaction contains three factors which combine
seamlessly in QED: the Coulomb repulsion, the spin of the electron,
and its statistics. However, the last two factors contribute
\emph{separately} to the \Mol{} formula; the first term in the
square brackets in (\ref{EQ-MOLLER}) is the Rutherford (Coulomb) term,
the second the exchange and the third the spin term. From a purely
phenomenological point of view, one may therefore consider modifying
the effects of these factors separately.

Modifying the Coulomb repulsion may be a step too radical; we consider
it to be currently unwarranted. (See also Sec.\ \ref{FREE-FREE}.)
Leaving the Coulomb interaction untouched would require modifying the
effects of exchange, or spin, or both.  Here the \Mol{} formula itself
offers a pointer: If the last term in the big square brackets in
(\ref{EQ-MOLLER}) is dropped, what remains is the formula for the
scattering of two spin-zero fermions.  If only the first term in the
square brackets in (\ref{EQ-MOLLER}) is kept, we obtain the
(relativistic) Rutherford formula for equal charges and masses. Could
it be that some (energy-dependent) fraction of the particles scatter
according to the \Mol{} formula, but the rest scatter as classical
particles, as spin-zero bosons (\ref{SCALAR-B}) or even as
\emph{spin-zero fermions}?

{
\begin{figure}[ht]
\hfil
{
\begin{minipage}[t]{.95\linewidth}
\vspace{2ex}
\hfil\includegraphics*[width=0.9\linewidth,keepaspectratio=true]{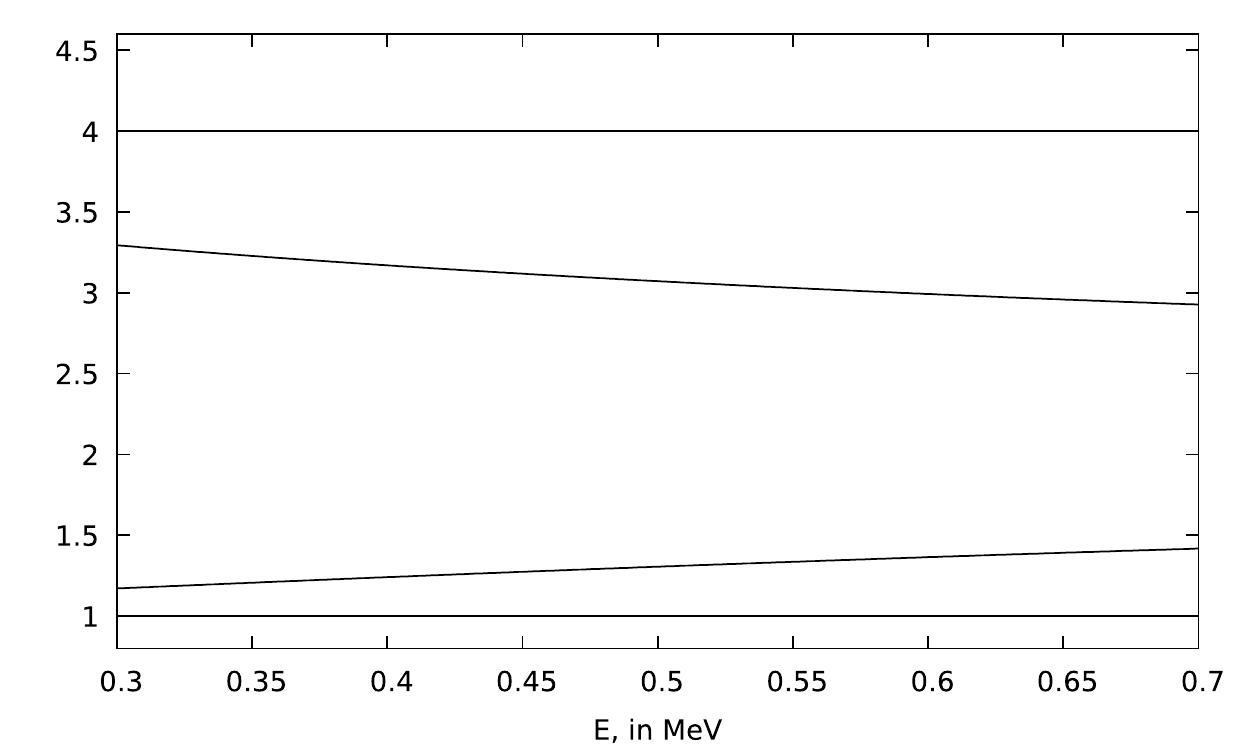}\hfil
\caption{From top to bottom: graphs of $R(E), B(E), M(E)$ and ${\sf
F}(E)$ for $x=0$. The four graphs are pairwise disjoint.}\label{FIG-CS}
\end{minipage}
}
\hfil
\end{figure}
}
  
Graphs of the terms in square brackets in the cross-sections for
(i)~Rutherford scattering $R(E)$, (ii)~the scattering of two spin-zero
bosons $B(E)$ (formula (\ref{SCALAR-B})), (iii)~\Mol{} scattering
$M(E)$, and (iv)~that of two spin-zero fermions~${\sf F}(E)$, between
$E=0.3$ and $0.7$ MeV at $x=0$, are shown in {Fig.}~\ref{FIG-CS}.
(The factor outside the square brackets is common to all four.)  One
sees immediately that the Rutherford and spin-zero boson
cross-sections are always significantly larger than the \Mol{}
cross-section.  Only the spin-zero fermions have a scattering
cross-section which is smaller than the \Mol{}! 
We therefore assume that

\vspace{1em}
\noindent{\bf Hypothesis I.} \emph{ An observed cross-section
$\sigma_{\sf obs}^{\prime}$ which is smaller than the \Mol\
cross-section may be represented as 
\begin{equation}\label{EQ-HYPO}
\xi\sigma_{\sf obs}^{\prime} = (1-\alpha)\,\sigma_{\sf M}^{\prime} +
        \alpha\,\sigma_{\sf F}^{\prime}
\end{equation}
where $\sigma_{\sf F}^{\prime}$ is the cross-section for the
scattering of two spin-zero fermions, $\alpha=\alpha(E) \in (0,1)$ is a
parameter which depends of $E$ but not on $x$, and $\xi=\xi(E)$ is a
small correction factor ($\xi(E)=1+\delta(E)$) which accounts for the
loss due to multiple scattering.
} 

\vspace{1em}\noindent Note that in the nonrelativistic limit

$$ \sigma_{\sf M}^{\prime} = \sigma_{\sf F}^{\prime}$$

\noindent so that $\sigma_{\sf obs}^{\prime}(E\rightarrow0)$ is
independent of $\alpha$.

\vspace{1em}\noindent{\bf Remarks II.} The right-hand side of
(\ref{EQ-HYPO}) can be written as the right-hand side of the \Mol{}
formula (\ref{EQ-MOLLER}), with the third term in square brackets
being replaced by
\begin{equation}\label{EQ-RMK-1}
(1-\alpha)\dfrac{(\gamma-1)^2}{4\gamma^2}\left(1+\dfrac{4}{1-x^2}\right)
\end{equation}
where $\alpha$ is the same as in (\ref{EQ-HYPO}). Alternatively, the
extra factor $(1-\alpha)$ can be absorbed in $\gamma$, as follows:
\begin{equation}\label{EQ-RMK-2}
\dfrac{(\gamma^{\prime}-1)^2}{4{\gamma^{\prime}}^2}
\left(1+\dfrac{4}{1-x^2}\right)
\end{equation}
where $\gamma^{\prime}(\gamma,\alpha) < \gamma$.  Neither of these
forms have as transparent an interpretation as (\ref{EQ-HYPO}); we shall
return to them briefly in {Sec.}~\ref{CONCLUSIONS}.

\vspace{1em}

We shall now make some numerical estimates concerning the feasibility
of subjecting the above hypothesis to experimental tests.

\section{Feasibility estimates}\label{ESTIMATES}

The essential part of the model defined by (\ref{EQ-HYPO}) is that $\xi$
and $\alpha$ depend on $E$ but not on $x$. For given $E$, measurement
of $\sigobs${} for two different values of $x$ will determine $\xi(E)$
and $\alpha(E)$ for that $E$; measurements of $\sigobs${} for the same
$E$ but other values of $x$ will then serve to test the hypothesis~I:
if the $\xi$ and $\alpha$ so determined fit the data for the same $E$
but other $x$, hypothesis~I may be accepted provisionally; if not, it
must be rejected.

To test (\ref{EQ-HYPO}) for given $E$, one has to measure
$\sigobs(x,E)$ for different values of $x$.  On the one hand, this
should be done for as many values $x$ as possible to accept or reject
the hypothesis~I with confidence; on the other hand, adjacent values
of $x$ for which $\sigobs(x,E)$ is measured should be separated enough
to distinguish between the corresponding values of $\sigobs(x,E)$.
These two requirements are in conflict, and the feasibility of the
experiment will depend on whether they can be met simultaneously
within the experimental error.

One more factor has to be taken into account. At small angles (those
utilised for transmission electron microscopy, TEM; see
\cite{EGERTON2011,EGERTON2016}) the \ee{} scattering cross-sections
are orders of magnitude greater than those at large angles (say for
$x=0$, as in the APW experiments). An experiment to detect possible
departures from the \Mol{} formula would have to be like the APW
experiments -- in which individual scattering events are detected and
counted -- and not like TEM runs in which images are formed by scattering
events too numerous to count; the scattering angles at which the
cross-sections are measured would have to be significantly larger than
those used in TEM.

Numerical calculations show that for $E$ from 0.3 to 0.7 MeV and $x$
from 0.10 to 0.40, the scattering angle $\theta$ lies between
$36.6^{\circ}$ and $25.0^{\circ}$. These angles should be far enough
from TEM regimes for individual events to be detectable.

Let $\sigobs(x,E)$ be measured at $x=x_0,\ldots x_n$, with $x_k < x_{k+1}$.
To simplify the notation, write
$$\spr_k(E)\quad\mathrm{for}\quad\sigobs(x_k,E)$$
Then, to distinguish  between $\spr_k(E)$ and
$\spr_{k+1}(E)$, the error in the measurement of
$\spr_k(E)$ (or $\spr_{k+1}(E)$) must be discernibly less
than 

\begin{equation}\label{DELTA-K}
\begin{array}{lcl}
\Delta_k(E) &=& 100\left(\dfrac{\xi\spr_{k+1}(E)
-\xi\spr_{k}(E)}{{\xi\spr_{k+1}(E)}}\right)\\[1.5em] 
&=& 100\left(1 - \dfrac{\spr_{k}(E)}{\spr_{k+1}(E)}\right)
\end{array}
\end{equation}

\vspace{1mm}\noindent
The factors $100$ in the above equation have been introduced to
express the $\Delta_k$ as percentages of $\spr_{k+1}$. Note that
$\Delta$ is independent of $\xi$.

For the energy values under consideration, $\spr_k(E)$ is an
increasing function of $k$, so that the $\Delta_k(E)$ are positive
numbers.  The numerical value of (\ref{DELTA-K}) places a limit on how
close $x_k$ can be to $x_{k+1}$ for $\sigobs(x_k,E)$ to be
distinguishable from $\sigobs(x_{k+1},E)$.

The errors in the APW experiments, carried out in the late 1940s and
early 1950s, were reported as 7--8\%.  (In their respective theses,
both Page \cite{PAGE1950} and Ashkin \cite{ASHKIN1952} go into
considerable detail about the various sources of error and their
individual contributions to the total error.)  However, in a Page-type
experiment the \emph{ratios} $\Delta_k$ can be determined
\emph{directly} from the observed count rates (cf.\ Ashkin's remarks
about the greater accuracy of the ratio $\rho(E,x)$ in
{Sec.}~\ref{ASHKIN}), and should be much more accurate than the
individual cross-sections. It will not be overly optimistic to assume
that the $\Delta_k$ would not exceed 4--5\% even with the technology
used by Page and Ashkin seventy years ago. We shall continue our
analysis on the assumption that it will be possible to distinguish
between $\sigobs(x_k,E)$ and $\sigobs(x_{k+1},E)$ if $\Delta_k > 5\%.$

We have calculated $\Delta_k$ numerically for 

\begin{enumerate}

\nitem $E=0.4, 0.5, 0.6\; \mbox{and}\;0.7$.

\nitem For each $E$,  $x_k$ = 0.10, 0.15, 0.20, 0.25, 
0.30 and 0.35.

\nitem For each pair $(E,x)$, $\alpha$ = 0.1, 0.2, 0.3, 0.4
and 0.5.  

\end{enumerate}
The results of the calculations are shown in tabular form in the
Appendix, one table for each value of $\alpha$. One sees from these
tables that $\Delta_k $ is less than $5.0$ only for a few values of
$E$ and $\alpha$ at $x=0.10$.  If the value $x=0.10$ is altogether
excluded from the measurements, then $\Delta_k\geq 6.46$, and one has
five sets of data points available: $x_k = 0.15, 0.20, 0.25, 0.30,
0.35$.  Any two of these will fix the values of $\xi$ and $\alpha$ in
(\ref{EQ-HYPO}) for given $E$; the other three can then be used to
test the hypothesis itself. If one excludes the data point $x_k=0.15$
as well, one will still have two data points for a rougher test of the
hypothesis at any $E$, with $\Delta_k \geq$ (a very accommodating)
$8.21$.

We also see from the tables that, for larger $x_k$, the $\Delta_k$ are
so large that more data points can be picked in the interval $x \in
[0.2,0.35]$ than the three we have chosen to determine the feasibility
of the experiment. Details are left to the experimentalist.

We may therefore conclude that hypothesis~I is indeed susceptible to
experimental test, even with a repetition of Page's experiment.


\subsection{Test for multiple scattering, II}\label{TMS-II}

Begin with the observation that for fixed $E$ and $x_1\neq x_2$, the
ratio of cross-sections 
\begin{equation}
\upsilon(E;x_1,x_2) = \dfrac{\spr(E,x_1)}{\spr(E,x_2)}
\end{equation}
can be determined more precisely than either cross-section if both
measurements are carried out with the same apparatus and the same
intensity of the incident beam. The ratio $\upsilon$ will simply be the
ratio of the observed count rates. This observation can be used
to turn the experiment described above into a test for multiple
scattering with very little extra effort. The extra effort would be to
measure $\spr(E,x)$ for each $E$ at $x=0$ as well, and then to
compute the ratios
\begin{equation}\label{OBS-U-RATIOS}
\Upsilon_{\sf obs}(E,x_k) = \dfrac{\spr(E,x_k)}{\spr(E,x=0)}
\end{equation}
for $x_k = 0.10, 0.15,\ldots 0.35$ from the observed count rates.  For
any given $E$, these points should be plotted together with the graph
of the calculated ratio of $\Upsilon_{\sf M}(E,x)$ versus $x$ for
$x\in [0,0.4]$.

As $x$ increases, the scattering angle will decrease, and so will the
path traversed through the foil by the scattered particles. If
multiple scattering is significant, then the $\Upsilon_{\sf obs}$ will
depart nore and more from the calculated graph with decreasing $x$.


\section{Free-free scattering; the experiment of Williams et
al}\label{FREE-FREE}

In 2014, Williams et al published the results of an important
experiment on the `Scattering of free electrons by free electrons',
which was probably the first of its kind \cite{WILLIAMS2014}. (In the
following, we shall call this \emph{free-free scattering.}) In this
experiment, the relative velocity $v/c$ of the electrons varied
between $10^{-2}$ and $10^{-1}$, i.e., $\gamma$ varied from $1.00005$
and $1.00504$, small enough for the collisions to be considered
nonrelativistic. As we have noted earlier, in the nonrelativistic
limit the scattering cross-section does not distinguish between two
electrons and two spin-zero fermions (of mass $m=m_e$), so that the
abovementioned experiment does not provide the information that we are
seeking. Page and Ashkin measured the cross-section at a single
scattering angle (for given $E$); Williams et al measured the
cross-sections at several different scattering angles; their results
agreed with the theory based on a strictly Coulomb central force.

Williams et al write that their experiment verified the \Mol\
formula to within \PM4\%. (One could say, with more drama but equal
justice, that their experiment showed that the particles scatter like
scalar fermions to within \PM4\%.) Their runs yielded 48 data points,
of which four were wildly off, and were disregarded. They added that
`The additional time required for better precision was not pursued
because the [energy] dependence was clear and consistent with
Rutherford scattering'. By `Rutherford scattering' they meant the
factor $\beta^{-4}$ \emph{outside} the square brackets in
(\ref{EQ-MOLLER}), which is common to all cross-sections under the
Coulomb force.  This factor also shows that, other things being equal,
at higher energies (relative velocities), longer runs needed to
achieve higher accuracies -- or indeed to make any observations at all
-- will have to be very much longer: to get the same number of data
points as at $E$ = 2.5 keV, the run will have to be 2,340 times as
long at 0.2 MeV and 7,940 times as long at 1.0 MeV!

The multiple scattering problem could exist in free-free scattering as
well, at high intensities and large (beam) cross-sections. If it does,
one may be able to minimize its effect by using beams that are almost
two-dimensional. The present author is unable to assess the
feasibility of this suggestion, or that of scaling up the experiment
of Williams et al to the energy range 0.4--0.7 MeV. (Williams et al
write about extending the experiment to \emph{lower} relative
velocities.)  


\section{Implications for the theory}\label{CONCLUSIONS}

Runs of experiments that are more extensive and precise than those
of Page and Ashkin may have three possible outcomes.

\begin{enumerate}

\nitem The results agree with the \Mol{} formula.

\nitem The results depart from the \Mol\ formula, but do not validate
hypothesis~I.

\nitem The results depart from the \Mol\ formula, and validate
hypothesis~I.

\end{enumerate}

\noindent In the first case, we would conclude that there is no need
to attempt a revision of QED; the expreiments of Page and Ashkin had
unexplained errors. In the second case, we may have to consider a
non-Coulomb central force, which would entail a study of the angular
distribution (as opposed to Page and Ashkin, who studied only the
scattering angle corresponding to $x=0$ for each $E$).  As stated
earlier (and reinforced by the results of Williams et al), we consider
this possibility to be very remote, and shall not say anything further
about it. It is the third case -- an abundance of observations conform
to hypothesis I -- that will be our main concern, but we first need to
dispose of a somewhat different scenario that may be suggested.

Usual proofs of the spin-statistics theorem are based on relativistic
invariance. Although there have been several attempts to establish the
result in a nonrelativistic setting (Sudarshan \cite{ECGS1968},
Balachandran et al \cite{BAL1993}, Berry and Robbins \cite{BR1997}),
quantum field theorists continue to believe that the theorem breaks
down in nonrelativistic physics. Could it be that the spin-statistics
connection breaks down because the dynamics becomes effectively
nonrelativistic before the nonrelativistic limit is reached?

The answer is \emph{no}! At the energies we are considering, the
\emph{kinematics} of a two-electron system will certainly be
relativistic (even at $E = 0.2$\,MeV, $\gamma = 1.3914$).  Recall now
that mass is a superselection rule in Galilei-invariant theories
(Bargmann's superselection rule \cite{VB1954}). Therefore the
\emph{in}- and \emph{out}- states in our experiment will belong to
different superselection sectors. This will frustrate any sttempt to
combine relativistic kinematics with nonrelativistic dynamics.

As pointed out in Remark II, eq.~(\ref{EQ-HYPO}), the defining equation
of hypothesis I, can also be cast in two other forms: the \Mol\
formula with a modified third term (in the square brackets), given by
either (\ref{EQ-RMK-1}) or (\ref{EQ-RMK-2}). However, we have not been
able to find any physical interpretation for these forms.

There remains one further possibility: modifying the \emph{second} term 
$$-\;\dfrac{3}{1-x^2}$$
in square brackets in the \Mol{} formula (\ref{EQ-MOLLER}). This term
is negative, and multiplying it by a factor greater than one will
lower the cross-section. It will also change the angular distribution.
The effects will be most pronounced near the nonrelativistic limit.
Williams et al did not find any such effect \cite{WILLIAMS2014}.

To sum up, verification of hypothesis I will create serious problems
for the spin-statistics connection for electrons that are not in bound
states, and more generally for quantum electrodynamics and quantum
field theories. It should also be noticed that the energy range in
which the effect is observed is below the threshhold of pair creation,
where the notion of quantized fields is not called upon; indeed,
\Mol{} derived his formula within the Dirac theory, using his own
relativistic generalization \cite{MOLLER1931} of a trick devised by
Bethe \cite{BETHE1930} (see also \cite{ROQUE1992}, page 200). This may
suggest that the whole notion of particles as field quanta -- one of
the most basic notions of theoretical physics today -- has to be
re-examined, the repercussions of which will be almost unimaginable.
For example: is it really necessary to quantize the gravitational
field?  What questions does string theory answer? What are the objects
of which quantum mechanics is a mechanics? Let alone a theory of
everything, will we have a theory of \emph{anything} at all?


\section{Possible violation of Pauli's principle in bound states}\label{VIP}

So far we have been considering the possible violation of exchange
symmetry in states of free particles, albeit based on the assumption
that the scatterer is also a free particle. The possible violation of
the Pauli principle in \emph{bound} states was first tested
experimentally in 1990, and is the subject of a major ongoing
collaboration. We shall describe it briefly in the following.  For the
background, we refer to the general discussion of exchange symmetries
from the theorist's point of view in Haag's book \cite{HAAG1993} and
the references quoted there.  (The term \emph{para\-statistics} seems
to have been introduced by Dell'Antonio, Greenberg and Sudarshan in
\cite{DGS1964}). We first describe an ansatz which produces a `small'
violation of the Pauli exclusion principle, one which has led to
experiments. 

In Fermi-Dirac statistics, the operator identity $(\adag_j)^2 = 0$
(which follows from the anticommutation rules) ensures that the state
$j$ is not occupied more than once. If, instead, we had the operator
identitiy $(\adag_j)^3=0$, it would imply that the state $j$ can be
occupied \emph{twice}, but no more. This, of course, would require
trilinear commutation relations.

In 1987, Ignatiev and Kuzman published a model with trilinear
commutation relations which contained a small parameter $\beta$
\cite{IK1987}. Their model, which had only one level, was defined by

\begin{equation}\label{IK-ANSATZ-1}
\begin{array}{rcrcc}
a^2\adag &+& \beta^2\adag a^2 &=& \beta^2a\\
a^2\adag &+& \beta^4\adag a^2 &=& \beta^2a\adag a\\
\end{array}
\end{equation}

\noindent and their hermitian conjugates, together with
\begin{equation}
a^3 =0,\quad (\adag)^3 =0
\end{equation}
This model has proved difficult to generalize to systems with many
degrees of freedom (for details, see \cite{IGNATIEV2005} and the
references cited there), but it has the advantage of being
experimentally testable. For example, if one of the two $1s$ states of
an atom can be doubly occupied, albeit briefly, then its existence may
be revealed by $2p\rightarrow1s$ transitions that would otherwise be
forbidden when both $1s$ states are occupied.

An experiment to test this possibility in an open system (to bypass
the Messiah-Greenberg superselection rule \cite{MG1964}) was performed
by Ramberg and Snow in 1990 \cite{RS1990}. They passed a large current
through a copper trough and looked for $2p\rightarrow 1s$ transitions.
Normally the $1s$ shell is already filled with two electrons, so that
the transition, of energy $\sim$ 8.05 keV, is forbidden. A small
violation of Pauli principle of the type (\ref{IK-ANSATZ-1}) would,
however, allow some such transitions. Owing to screening by existing
electrons, these transitions would have an energy $\sim$ 7.7 keV, and
the resulting X-rays can be detected. Ramberg and Snow found no such
transitions during their run, which gave the bound $\beta^2 \leq
1.7\times10^{-26}$.

The Ramberg-Snow experiment was greatly refined by a large
international collaboration (the VIP collaboration) at the Laboratoria
Nazionale di Frascati (LNF) of the INFN, Italy \cite{VIP2006}. They
obtained a much improved lower bound of $\beta^2/2\leq
4.5\times10^{-28}$. This experiment was subsequently moved the
Laboratoria Nazionale di Gran Sasso (LNGS) under the Gran Sasso
mountain in the Abruzzi. In its new form (which is constantly being
refined) it is known as the VIP 2 experiment \cite{PICHLER2016}. The
latest published bound from this experiment is
$\beta^2/2\leq3.4\times10^{-29}$ \cite{SHI2018}.

The project is continuing. For a status report from April 2019, see
\cite{PC2019}. A good discussion of the theoretical background
and derivation of the bound may be found in \cite{MILOTTI2018}.


\vspace{2em}

\nin{\bf Acknowledgements}

\vspace{1em}\noindent The author would like to thank Dr Luca De Paolis
for bringing the VIP and VIP 2 experiments to his attention, and Dr
Mayer Goldberg for help with the numerical calculations and the
graphs.


\newpage


\setcounter{table}{0}
\renewcommand{\thetable}{A\arabic{table}}

\section*{Appendix: Tables of $\Delta_k(E,x;\alpha)$}\label{TABLES}

\vspace*{1em}
\noindent{\bf Note}: Values of $\Delta_k$ \emph{smaller than} 5.0 are
shown in {\bf boldface}.

\vspace*{3em}

\small

\newcolumntype{d}[0]{D{.}{.}{2}}

\begin{table}[h]
\begin{center}
\begin{tabular}{|c|d|d|d|d|}\hline
    &\multicolumn{4}{c|}{$E$, in MeV}\\ \cline{2-5}
\raisebox{1em}{${x}$}   & 0.4  & 0.5  & 0.6  & 0.7\\[2mm]\hline
0.10& 5.12 & \bf4.\bf{95} &\bf4.\bf{80} & \bf4.\bf{68}\\
0.15& 7.04 & 6.81 & 6.62 & 6.46\\
0.20& 8.87 & 8.61 & 8.39 & 8.21\\
0.25&10.64 &10.36 &10.12 & 9.91\\
0.30&12.36 &12.07 &11.81 &11.60\\ 
0.35&14.07 &13.78 &13.53 &13.31\\ \hline
\end{tabular}
\caption{$\Delta_k(E)$ for $\alpha=0.1$}
\label{TABLE-1}
\end{center}
\end{table}
\begin{table}[h]
\begin{center}
\begin{tabular}{|c|d|d|d|d|}\hline
    &\multicolumn{4}{c|}{$E$, in MeV}\\ \cline{2-5}
\raisebox{1em}{${x}$}   & 0.4  & 0.5  & 0.6  & 0.7\\[2mm]\hline
0.10& 5.20 & 5.04 & \bf4.\bf90 & \bf4.\bf68\\
0.15& 7.14 & 6.93 & 6.75 & 6.60\\
0.20& 8.99 & 8.75 & 8.54 & 8.37\\
0.25&10.77 &10.50 &10.28 & 10.09\\
0.30&12.49 &12.22 &11.99 &11.79\\ 
0.35&14.18 &13.93 &13.70 &13.50\\ \hline
\end{tabular}
\caption{$\Delta_k(E)$ for $\alpha=0.2$}
\label{TABLE-2}
\end{center}
\end{table}\protect{\nopagebreak}
\begin{table}[h]
\begin{center}
\begin{tabular}{|c|d|d|d|d|}\hline
    &\multicolumn{4}{c|}{$E$, in MeV}\\ \cline{2-5}
\raisebox{1em}{${x}$}   & 0.4  & 0.5  & 0.6  & 0.7\\[2mm]\hline
0.10& 5.28 & 5.13 & 5.01 & \bf4.\bf90\\
0.15& 7.24 & 7.05 & 6.89 & 6.75\\
0.20& 9.10 & 8.89 & 8.70 & 8.54\\
0.25&10.89 &10.66 &10.45 & 10.09\\
0.30&12.62 &12.38 &12.17 &11.99\\ 
0.35&14.33 &14.09 &13.88 &13.70\\ \hline
\end{tabular}
\caption{$\Delta_k(E)$ for $\alpha=0.3$}
\label{TABLE-3}
\end{center}
\end{table}
\begin{table}[h]
\begin{center}
\begin{tabular}{|c|d|d|d|d|}\hline
    &\multicolumn{4}{c|}{$E$, in MeV}\\ \cline{2-5}
\raisebox{1em}{${x}$}   & 0.4  & 0.5  & 0.6  & 0.7\\[2mm]\hline
0.10& 5.36 & 5.23 & 5.12 & 5.02\\
0.15& 7.34 & 7.18 & 7.03 & 6.91\\
0.20& 9.23 & 9.03 & 8.87 & 8.72\\
0.25&11.02 &10.81 &10.63 & 10.28\\
0.30&12.76 &12.54 &12.35 & 12.19\\ 
0.35&14.46 &14.25 &14.06 & 13.90\\ \hline
\end{tabular}
\caption{$\Delta_k(E)$ for $\alpha=0.4$}
\label{TABLE-4}
\end{center}
\end{table}
\begin{table}[h]
\begin{center}
\begin{tabular}{|c|d|d|d|d|}\hline
    &\multicolumn{4}{c|}{$E$, in MeV}\\ \cline{2-5}
\raisebox{1em}{${x}$} 
    & 0.4  & 0.5  & 0.6  & 0.7\\[2mm]\hline
0.10& 5.45 & 5.33 & 5.23 & 5.15\\
0.15& 7.45 & 7.31 & 7.18 & 7.07\\
0.20& 9.35 & 9.19 & 9.04 & 8.91\\
0.25&11.16 &10.98 &10.82 &10.68\\
0.30&12.89 &12.71 &12.55 &12.41\\ 
0.35&14.60 &14.42 &14.26 &14.12\\  \hline
\end{tabular}
\end{center}
\caption{$\Delta_k(E)$ for $\alpha=0.5$}
\label{TABLE-5}
\end{table}
\vfill*\pagebreak\pagebreak

\vfill
\end{document}